\documentclass[useAMS,usenatbib]{mn2e}
\usepackage{graphicx}
\usepackage{lscape}
\usepackage{ulem}

\def\kms{km~s$^{-1}$}
\def\cm{cm$^{-2}$}
\def\lya{Ly$\alpha$}
\def\nhi{$N$(H~I)}
\def\hkpc{$h_{70}^{-1}$ kpc}
\def\dind{$D$-index}

\title[Pre-selection of damped Lyman alpha systems]
{An efficient technique for pre-selecting low redshift damped Lyman 
alpha systems.}
\author[S. L. Ellison]
{Sara L. Ellison$^1$\thanks{Email: sarae@uvic.ca}
\\
$^1$Dept. Physics \& Astronomy, University of Victoria, 3800 Finnerty Rd, 
Victoria, BC, V8P 1A1, Canada\\
}

\begin{document}
\maketitle

\begin{abstract}

The number of  $z \sim 1$ damped Lyman alpha systems (DLAs, log \nhi\ $\ge$ 
20.3) per unit 
redshift is approximately 0.1, making them relatively rare objects.
Large, blind QSO surveys for low redshift DLAs are therefore an expensive 
prospect for space-borne UV telescopes.  Increasing the efficiency
of these surveys by pre-selecting DLA candidates
based on the equivalent widths of metal absorption lines has
previously been a successful strategy.  However, the success 
rate of DLA identification is still only $\sim$ 35\% when simple equivalent 
width cut-offs are applied, the majority of systems having $19.0 <$ log
\nhi\ $<20.3$.
Here we propose a new way to pre-select DLA candidates.  
Our technique requires
high-to-moderate resolution spectroscopy of the Mg~II $\lambda$ 2796
transition, which is easily accessible from the ground for $0.2 \sol z
\sol 2.4$.  We define the \dind, the ratio of the line's equivalent
width to velocity spread and measure this quantity for 19 DLAs and
8 sub-DLAs in archival spectra obtained with echelle spectrographs.
For the majority of absorbers, there is a clear distinction between
the \dind\ of DLAs compared with sub-DLAs (Kolmogorov-Smirnov 
probability = 0.8 \%).  Based on this pilot data sample, we find
that the \dind\ can select DLAs with a success rate of up to 90\%, 
an increase in
selection efficiency by a factor of 2.5 compared with a simple
equivalent width cut.  We test the
applicability of the \dind\ at lower resolution and find that it
remains a good discriminant of DLAs for FWHM $\sol$ 1.5 \AA.
However, the recommended \dind\ cut-off between DLAs and sub-DLAs
decreases with poorer resolution and we tabulate the appropriate
\dind\ values that should be used with spectra of different resolutions.

\end{abstract}


\section{Introduction}

The compilation of large spectroscopic datasets, such as the Sloan
Digital Sky Survey (SDSS), has proven a major boon for the study of QSO 
absorption lines.  These surveys have yielded many hundreds of
damped Lyman alpha systems (DLAs, absorption systems with \nhi\ $ \ge2 
\times 10^{20}$ \cm) which can be used to study
the evolution of the galactic neutral gas content at $z>1.6$ (e.g.
Prochaska, Herbert-Fort \& Wolfe 2005).  In the local universe,
several wide field 21cm surveys have characterised the $z=0$ distribution
of neutral hydrogen gas (e.g. Zwaan et al. 2005; Rosenberg \&
Schneider 2003; Ryan-Weber, Webster \& Staveley-Smith 2003).  
Bridging these extremes, over a redshift range
that is emerging as a crucial time in galaxy formation,
is a high priority for absorption line studies.  In practice,
however, the low redshift universe has been challenging to survey.
The \lya\ line is shifted into the UV at $z<1.5$ where space telescopes
are required.  Given the limited resources, large-scale,
blind surveys for low redshift DLAs are expensive and largely unpalatable
to time assignment committees.  For this reason, pre-selection of
low $z$ DLA candidates based on low resolution, ground-based spectra has
become an important first step in UV surveys (Rao \& Turnshek 2000;
Rao, Turnshek \& Nestor 2006).  The equivalent widths (EWs) of resonance
lines with red rest wavelengths, such as Mg~II $\lambda \lambda$ 2796, 2803
and Fe~II $\lambda$ 2600, are easily measured from the ground in
systems with \nhi\ $\sog 10^{19}$ \cm\ and $0.2 \sol z \sol 2.4$.
Rao et al. (2006) showed that $\sim$ 35\% of absorbers with rest frame  
EW(Mg~II $\lambda$ 2796, Fe~II $\lambda$ 2600) $>$ 0.5 \AA\
have column densities above the canonical DLA limit, \nhi $\ge 2 \times
10^{20}$ \cm.  Although absorbers below this limit (sub-DLAs)
make a non-negligible contribution to the total neutral gas
content of the universe at high redshifts (P\'eroux et al. 2005;
Prochaska et al. 2005), they are usually excluded from statistical
samples.  Therefore, although EW pre-selection has been a pivotal
strategem of recent Hubble Space Telescope (HST) surveys, the 35\%
`success rate' of metal line pre-selection could be considered low.
Moreover, whilst a confirmation
probability can be applied in a statistical sense, there is
no way of knowing which individual Mg~II absorber is most likely
to be a DLA, since there is no correlation between Mg~II EW
and \nhi.  For example, a DLA confirmation rate of 35\% applied to the
$0.6 < z < 1.7$ Mg~II system survey of Ellison et al. (2004a)
yields a DLA number density of $n(z)=0.11^{+0.08}_{-0.05}$ 
(Ellison 2005).  However, UV follow-up
of the 14 high EW Mg~II systems in the Ellison et al. sample
is still required in order to confirm the identity of the
expected 5 DLAs.

In this paper, we investigate whether ground-based spectra of metal 
lines at high resolution can provide additional
information to help discriminate between DLAs and sub-DLAs.
Our objective is to identify a quantitative measure that
will efficiently separate DLAs and sub-DLAs with a success
rate $>>$ 35\%.  If the inclusion of sub-DLAs 
(i.e. false positives) can be minimised, the obvious benefit
will be that UV follow-up for DLA confirmation and \nhi\
determination will be far more efficient. 
Moreover, if the pre-selection is complete for high \nhi\ DLAs,
any quantity weighted by \nhi, such as the total neutral
gas content, will be robust, even if a few lower \nhi\ systems
are missed in the pre-selection process.    
Finally, a high success rate DLA discriminant could be
applied to surveys of Mg~II absorption systems to directly infer 
the DLA number density to a relatively high degree of accuracy,
with no need for UV observations.

\section{Data}

In order to investigate the properties of DLAs and sub-DLAs,
we compile a sample of absorbers with known \nhi\ column
densities and echelle spectroscopy which covers the Mg~II
transition.    In practice, the high resolution spectroscopy
was obtained with either HIRES on Keck or UVES on the Very Large
Telescope (VLT),
see Table \ref{archive} for references.  In the final column of
Table \ref{archive} we give the FWHM resolution of the spectrum
at the wavelength of the redshifted Mg~II transition.
These spectra were generously provided by a number
of individuals (see Acknowledgements) with the continua already
normalized.  We return to the impact of continuum fitting
in the next section.

The first DLA surveys relied on an equivalent width selection based
on relatively low ($\sim$ 10 \AA\ FWHM) spectra.  This selection
was therefore susceptible to blending of absorbers below the
DLA \nhi\ limit.  As shown by Ellison (2000), for EW $<$ 10 \AA,
the fraction of true DLAs is $<<$ 50\%.  We therefore only
select absorbers from the literature whose \nhi\ is based on spectra 
with a FWHM
resolution of at least 5 \AA.  In most cases, this is sufficient
to adequately fit damping wings with errors typically $\le \pm$ 0.1
dex (e.g. Ellison et al 2001; Rao, Turnshek \& Nestor 2006).
In Table \ref{archive} we present the QSOs used in this
study, as well as references to the \nhi\ column densities and
metal abundances which have been normalised to the solar (meteoritic)
scale of Lodders (2003).  We have assumed that ionization corrections
to Zn~II and Fe~II are negligible, which has largely been found to
be the case even in lower column density systems (P\'eroux et al.
2003).

\begin{center}
\begin{table*}
\caption{Archival QSO Data}
\begin{tabular}{lcccccccc}
\hline
QSO & $z_{\rm abs}$ & \nhi\ & [Zn/H] & [Fe/H] & S/N & $D$ & Refs (\nhi,[X/H])
& Resolution (\AA) \\
\hline
 Q0058+019 & 0.613 & 20.08$\pm$0.20  & $+0.10$ & $-0.31$ & 65 & 8.5 &  1,1 & 0.10\\
 Q0100+130 & 2.309 & 21.37$\pm$0.08  & $-1.53$ & $-1.76$ & 35 & 7.4 & 2,2  &0.20\\
Q0512$-$333A & 0.931 & 20.49$\pm$0.08  & ... & $-1.49$ & 60 & 6.2 & 3,3 & 0.17\\
Q0512$-$333B & 0.931 & 20.47$\pm$0.08  & ... & $>-1.29$ & 45 & 6.5 & 3,3  & 0.17\\
 Q0827+24  & 0.525 & 20.3$\pm$0.05   & $<-0.04$ & $-1.02$ & 20 & 8.5 & 4,5 & 0.08\\
 Q0841+129 & 2.375 & 20.99$\pm$0.08  & $-1.52$ & $-1.70$ & 17 &  8.7 & 6,6 & 0.21\\
Q0957+561A & 1.391 & 20.3$\pm$0.1   & $<-0.75$ & $-1.31$ & 30 & 7.6 & 7,7 & 0.12\\
Q0957+561B & 1.391 & 19.90$\pm$0.10  & $<-0.31$ & $-1.03$ & 25 & 8.2 & 7,7 & 0.12\\
 Q1101$-$264 & 1.838 & 19.50$\pm$0.05  & $<-0.56$ & $-1.46$ & 40 & 5.8 & 8,8 & 0.17\\
 Q1104$-$180 & 1.662 & 20.85$\pm$0.01  & $-1.00$ & $-1.55$ & 40 &  9.0 & 9,9  & 0.15\\
 Q1122$-$168 & 0.682 & 20.45$\pm$0.05  & $<-1.32$ & $-1.32$ & 45 & 8.2 & 10,10 & 0.11\\
 Q1151+068 & 1.774 & 21.30$\pm$0.08   & ... & ... & 15 & 8.6 & 11  & 0.18\\
 Q1157+014 & 1.944 & 21.60$\pm$0.10  & $-1.24$ & $-1.61$ & 30 & 8.2 & 6,6  & 0.23\\
 Q1210+173 & 1.892 & 20.63$\pm$0.08  & $-0.86$ & $-1.09$ & 20 & 6.9 & 6,6  & 0.17\\ 
 Q1223+175 & 2.466 & 21.44$\pm$0.08  & $-1.68$ & $-1.69$ & 15 & 8.4 & 11,11 & 0.23\\
 Q1247+267 & 1.223 & 19.88$\pm$0.10  & $-1.02$ & $-1.37$ & 75 & 5.0 &  12,12 & 0.12\\
 Q1331+170 & 1.776 & 21.14$\pm$0.08  & $-1.23$ & $-1.99$ & 45 & 6.5 & 2,2  & 0.18\\
 Q1351+318 & 1.149 & 20.23$\pm$0.10  & $-0.27$ & $-0.96$ & 30 & 6.3 & 12,12 & 0.12\\
 Q1451+123 & 2.255 & 20.30$\pm$0.15  & $-1.08$ & $-1.44$ & 6 & 8.3 & 8,8 & 0.16\\
 Q1622+238 & 0.656 & 20.4$\pm$0.1   & ... & ... & 27 & 8.5 & 4 & 0.12\\
 Q1629+120 & 0.900 & 19.70$\pm$0.04  & $-0.18$ & $-0.86$ & 33 & 5.7 & 13,14  & 0.12\\
 Q2128$-$123 & 0.430 & 19.37$\pm$0.08  & ... & $>-0.74$ & 28 & 5.8 & 10,10 & 0.09\\
 Q2206$-$199 & 1.920 & 20.44$\pm$0.08  & ... & $-2.54$ & 15 & 8.1 & 15,15  & ... \\
 Q2230+023 & 1.859 & 20.00$\pm$0.10  & ... & ... & 20 & 5.4 & 6  & 0.17\\
  Q2231$-$001 & 2.066 & 20.53$\pm$0.08  & $-0.86$ & $-1.18$ & 40 & 7.8 & 2,2  & 0.20\\
 Q2343+125 & 2.431 & 20.35$\pm$0.05  & $-0.74$ & $-1.26$ & 15 & 8.2 & 2,2  & 0.21\\
Q2348$-$144 & 2.279 & 20.59$\pm$0.08  & $<-1.94$ & $-2.22$ & 35 & 9.2 & 6,6  & 0.20\\
\hline 
\end{tabular}\label{archive}
\\ Upper limits are 3$\sigma$. 1 - Pettini et al. (2000);  2 - Dessauges-Zavadsky et 
al. (2004); 3 - Lopez et al. (2005); 4 - Rao \& Turnshek (2000); 
5 - Kulkarni et al. (2005); 6 - Dessauges-Zavadsky et al. (2006);
7 - Churchill et al. (2003); 8 - Dessauges-Zavadsky et al. (2003); 9 - 
Lopez et al (1999); 10 - Ledoux, Bergeron \& Petitjean (2002); 
11 - Dessauges-Zavadsky, private communication; 12 - Pettini et al.
(1999);  13 - Rao, Turnshek \& Nestor (2006); 14 - This work; 
15 - Pettini et al. (2002);  
\end{table*}
\end{center}

In addition to the referenced archival data, we present previously
unpublished data for the $z_{\rm abs} = 0.900$ sub-DLA towards
 Q1629+120 whose \nhi\ was determined by Rao et al. (2006).  
The echelle data were obtained as part of ESO program
69.A-0410(A) (PI, Athreya) on the UVES spectrograph on UT2 at the VLT.
A total of 12,000 seconds of data were obtained in service mode
between April and May 2002 using Dic1 346--580, yielding a wavelength
coverage of 3100--3870, 4800--5750, 5850--6800 \AA.  
Data reduction was executed using the
MIDAS routines of the UVES pipeline as described in full by
Ellison, Ryan \& Prochaska (2001).  We fitted the continuum using
a 3rd order Chebyshev polynomial based on absorption-free regions
$\sim$ 10 \AA\ wide either side of the Mg~II absorption.
The final `archival' sample used in this study therefore
consists of 19 DLAs and 8 sub-DLAs which have all been observed
with either UVES or HIRES and therefore have typical FWHM resolutions
of 6 -- 8 \kms.

\section{The $D$-index}

\begin{figure}
\centerline{\rotatebox{0}{\resizebox{9cm}{!}
{\includegraphics{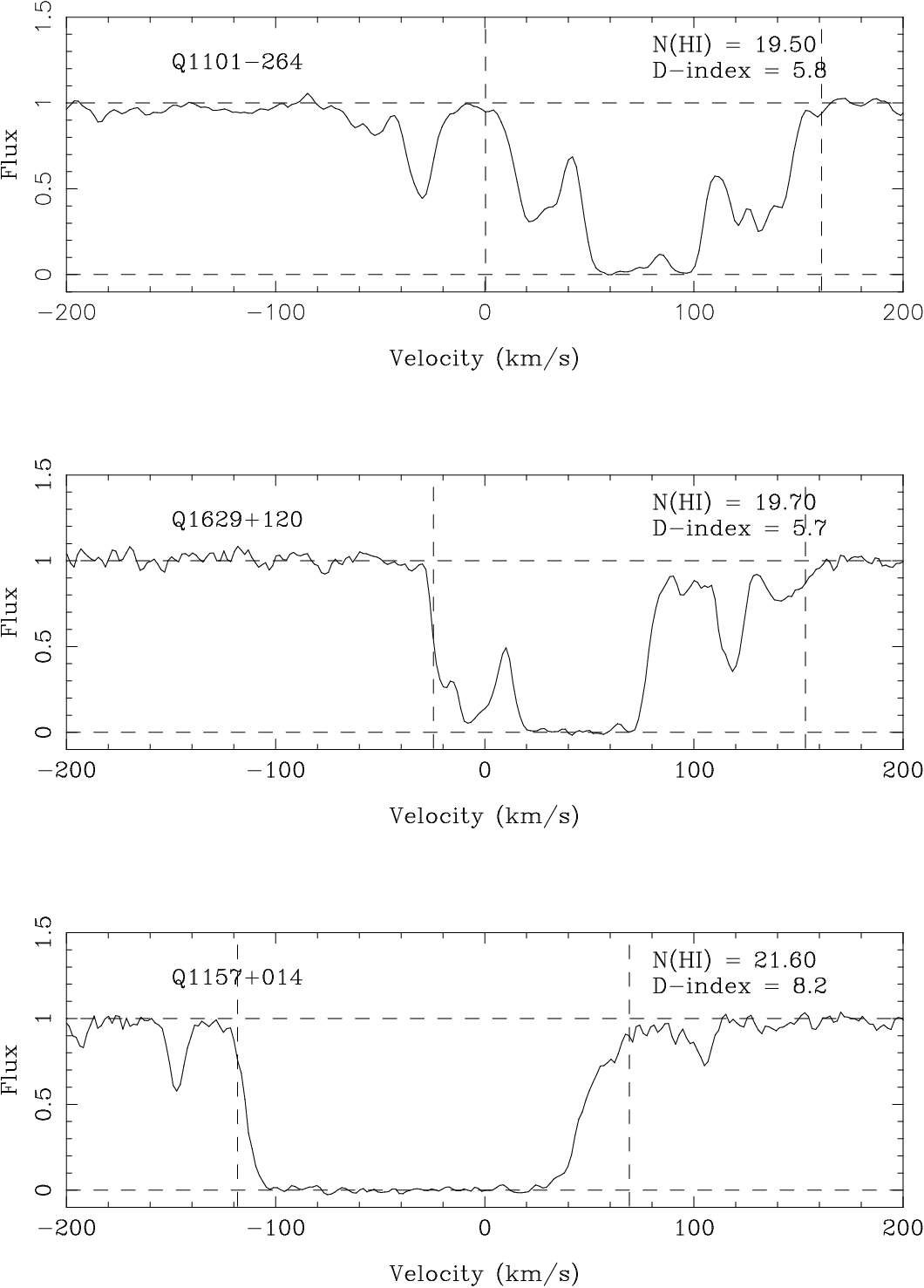}}}}
\caption{\label{3spectra} Examples of Mg~II line profiles for low 
(upper and middle panels) and high
(lower panel) $D$-indices.  Qualitatively, a low \dind\ means more
residual flux between the limits of the velocity spread (as shown
by the vertical dashed lines). }
\end{figure}

From the high resolution data, we measure the rest frame EWs
and velocity spread of the Mg~II $\lambda$ 2796 absorption
line.  In calculating these quantities, we exclude `detached'
absorption components where the continuum is recovered
and only include the complex with the
largest EW.  We set $\lambda_{\rm min, max}$ as the limits of
absorption that are significant at the 3 $\sigma$ level below
the continuum, see Figure \ref{3spectra}. 
In this study we restrict
ourselves to the Mg~II $\lambda$ 2796 line since we are working
with archival data with restricted wavelength access.  Nonetheless,
Mg~II is a useful line since it has a high oscillator strength
(and is therefore easy to detect) and is observable for most
low-to-intermediate redshift DLAs.
 
\begin{figure}
\centerline{\rotatebox{270}{\resizebox{6cm}{!}
{\includegraphics{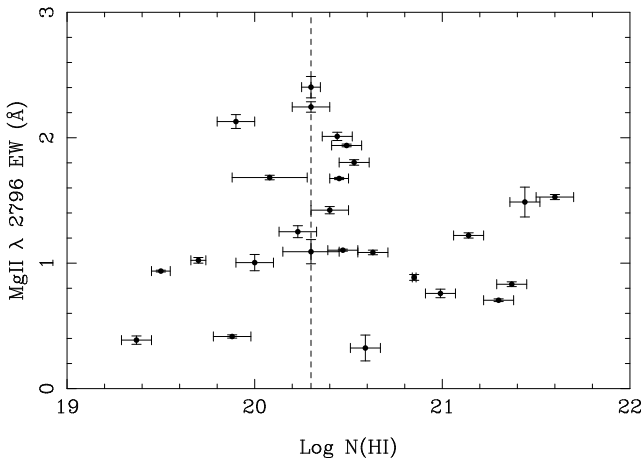}}}}
\caption{\label{ew_vs_hi} The rest frame equivalent widths (EW)
versus the \nhi\ of the archival absorbers listed in Table \ref{archive}.
There is no correlation between column density and Mg~II EW.}
\end{figure}

In Figure \ref{ew_vs_hi} we demonstrate the well known  scatter
between \nhi\ and Mg~II EW.  In our sample, the highest EWs are
exhibited by absorbers on the cusp of DLA classification.  
Similarly, moderate EWs $\sim$ 1 \AA\ can be associated
with absorbers whose \nhi\ column density ranges over two orders
of magnitude.  It has been previously noted (e.g. Nestor et al. 2003)
that the EW of Mg~II $\lambda$ 2796 is more an indication of
velocity spread than it is of \nhi.  This is the reason that
a simple EW cut-off applied to low resolution spectra 
contains such a high fraction of sub-DLAs.  An EW cut is unable to
distinguish between a few highly saturated Mg~II absorption components
and many lower column density clouds; the distinction between
these two is shown in Figure \ref{3spectra}.  Qualitatively, DLAs
tend to be characterised by the former, with wide, saturated
absorption troughs and sub-DLAs by the latter.


\begin{figure}
\centerline{\rotatebox{0}{\resizebox{9cm}{!}
{\includegraphics{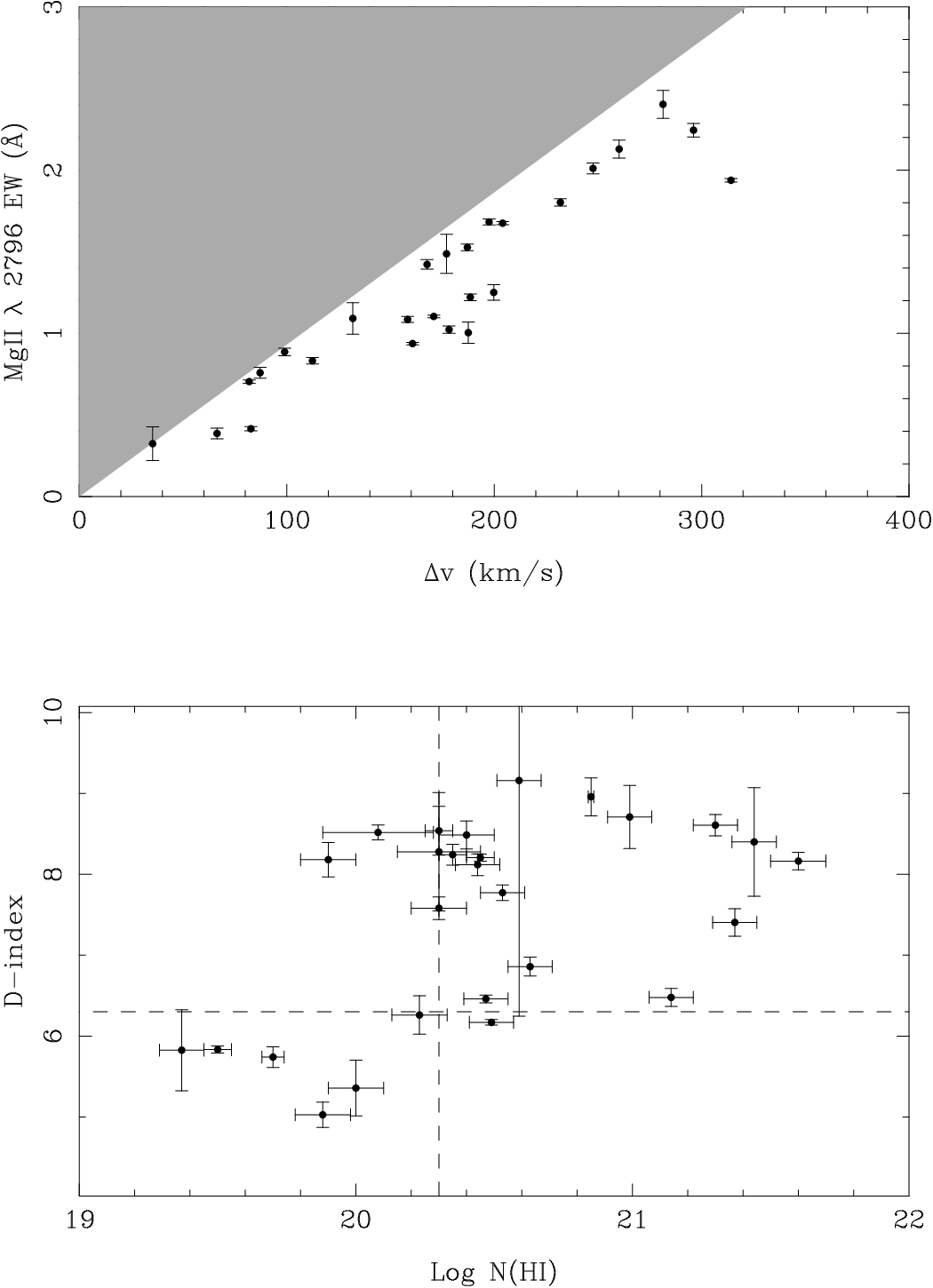}}}}
\caption{\label{dindex} Upper panel:  The EW of the Mg~II $\lambda$ 2796 
transition
versus the velocity spread, $\Delta v$.  The shaded region represents
forbidden values of EW for a given $\Delta v$.
Lower panel: $D = EW/\Delta v \times 1000$ as a function of \nhi.  
The horizontal
dashed line shows a D = 6.3 which most
effectively distinguishes the DLAs from lower column density absorbers.}
\end{figure}

In the upper panel of Figure \ref{dindex} we plot the EW (\AA) of 
Mg~II $\lambda$ 2796 versus its velocity spread ($\Delta v$, km/s).  
Within the bounds of $\lambda_{\rm min}$ and $\lambda_{\rm max}$,
which define the velocity spread, it is simple to show that
there is a maximum EW.  For this reason, there is a sharp cut-off
in EW in the upper panel of Figure \ref{dindex}, where the shaded 
area shows the locus of forbidden values.  In the lower panel
of Figure \ref{dindex} the data are plotted as a function of
\nhi\ and we combine the velocity and EW information in a single
parameter, $D$, where 

\begin{equation}
D = \frac{EW (\AA)}{\Delta v (km/s)} \times 1000 
\end{equation}

The error in the EW ($\delta$ EW)
is calculated from the 1 $\sigma$ spectral error arrays on a pixel-by-pixel
basis and propagated across the velocity width, $\Delta v$, of
the line.  Since we have explicitly defined $\Delta v$, the error
in the \dind\ ($\delta$ D) is simply calculated from 

\begin{equation}\label{deqn}
\frac{\delta D}{D} = \frac{\delta EW}{EW}
\end{equation}

The errors do not include systematic uncertainties associated with,
for example, continuum fitting.  The effect of continuum fitting
should be negligible in the determination of the \dind\ for several
reasons.  First, the Mg~II transition usually lies well redwards of
the Ly$\alpha$ forest in a spectral region where the continuum is
relatively flat and unabsorbed.  Continuum fitting in these regions
is usually straightforward.  Second, since the majority of
the Mg~II lines are heavily saturated, they will be relatively
insensitive to errors in the continuum fit.  We experimented by
re-fitting the continuum of Q1629+120 with 5 different combinations
of polynomial and spectral region.  Re-calculating the \dind\
shows errors $\le$ 1\%, i.e. smaller than the random
errors derived from Equation \ref{deqn}.
We conclude that random errors in EW are likely to dominate over
systematics such as the continuum fit.

In Table \ref{archive} we list the $D$-indices for the archival sample.
The high \dind\ absorbers are those that are closest to the
shaded area in the upper panel of Figure \ref{dindex}, whilst
the low \dind\ absorbers have lower EWs for their $\Delta v$.
Returning to the example absorbers shown in Figure \ref{3spectra}
gives a visual impression of how the \dind\ relates to
spectral morphology.

There is a clear tendency for DLAs to have higher $D$-indices than 
sub-DLAs (lower panel, Figure \ref{dindex}).  Although there is not a 
clear dividing line between
the populations, a \dind\ cut-off of 6.3  includes 
all but 1 of the 19 DLAs (95\%).  We consider the fraction of
DLAs above the \dind\ cut-off to be the `success-rate' of this
pre-selection; for  $ D = 6.3$ that rate is 90\% in our archival
sample of 27 absorbers.  
A Kolmogorov-Smirnov (KS) test applied to the
$D$-indices shows that the DLAs and sub-DLAs have
only a 0.8\% probability of being drawn from the same population. 
In the limited sample available to us, we
also note that all DLAs with log \nhi\ $>$ 20.5 are included
by a $D > 6.3$ criterion.  We experimented with
including metallicity information into the \dind, both
the [Zn/H] abundance which is generally the preferred metallicity
indicator, but also [Fe/H].  The abundance values used in our
testing are given in Table \ref{archive}.  We found that
including abundance (absolute or relative) information
did not improve the observed distinction between DLAs and sub-DLAs.
Although it is desirable
to populate this parameter space with many more systems, the
\dind\ appears to be a promising method for pre-selecting DLAs
if high resolution data are available.

\section{Data Requirements}

\subsection{Resolution}

We now consider the resolution requirements for the application
of the \dind.  We test this by convolving
each of the spectra used in this study with a Gaussian of 
steadily increasing FWHM (applicable at the wavelength of
Mg~II $\lambda$ 2796) to simulate
the effect of decreasing spectral resolution.  We re-calculate
the effective minimum and maximum wavelengths (which
are used in the determination of the EW and $\Delta v$)
taking into account the smoothing of the data on the S/N
ratio.  The \dind\ of each
convolved absorber and the KS probability of the two populations
are then re-calculated.

\begin{figure}
\centerline{\rotatebox{0}{\resizebox{9cm}{!}
{\includegraphics{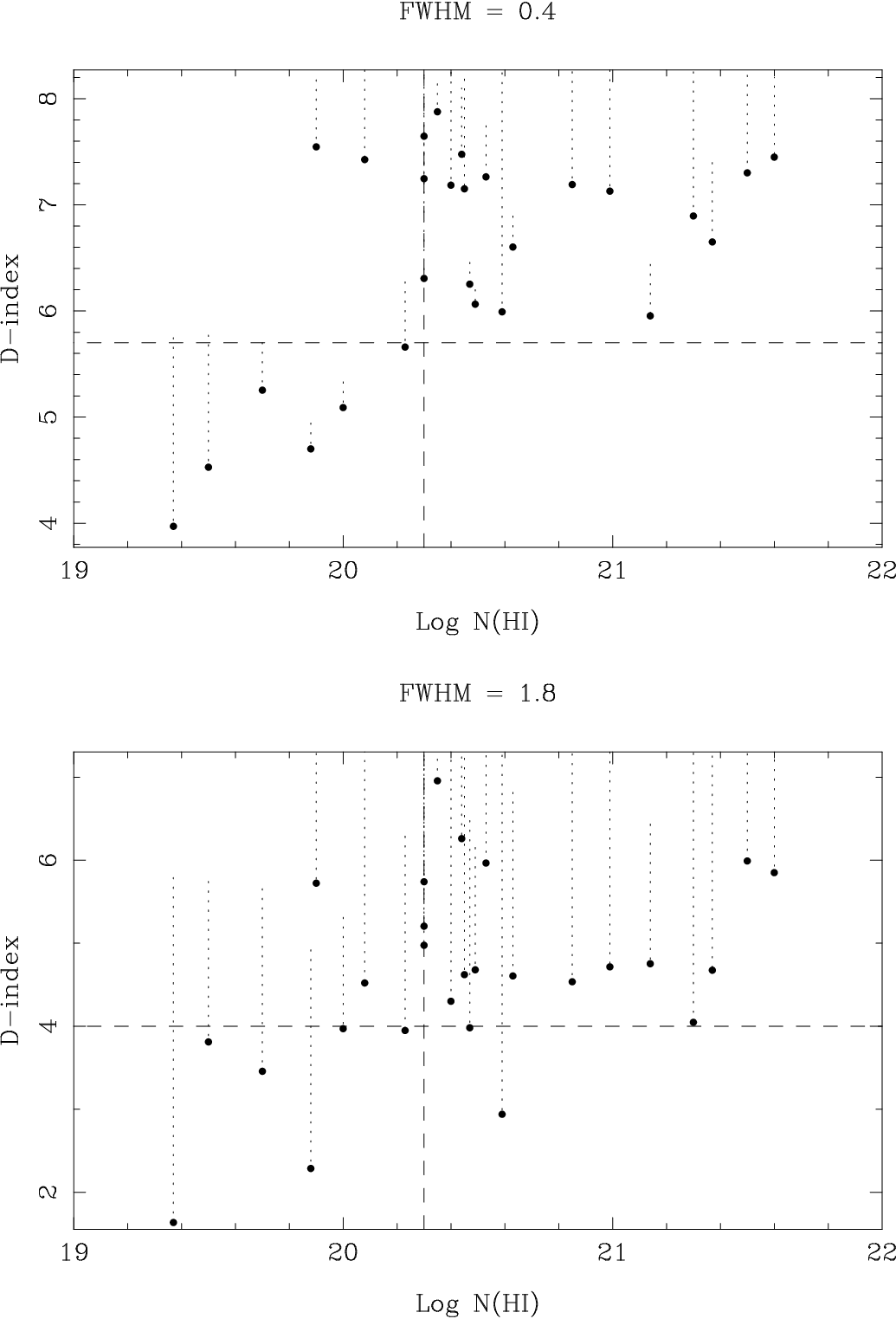}}}}
\caption{\label{convolve} Each of the spectra in our archival sample
is convolved with a Gaussian of FWHM shown above each panel (see
Table \ref{convolve_table} for the results of the full test).
Poorer resolution decreases the recommended \dind\ cut-off
for distinction between DLAs and sub-DLAs.  For $FWHM > 1.5$
the \dind\ no longer provides a robust discriminant of the two
populations.  The dotted lines show the vector shift between the
original \dind\ (unconvolved spectrum) and the convolved data.}
\end{figure}

We find that the DLA and sub-DLA populations are well separated
for FWHM $<$ 1.5 \AA\ (at the wavelength of redshifted
Mg~II $\lambda$ 2796), as shown by the examples in Figure \ref{convolve}.  
However, the appropriate cut-off in \dind\ decreases with decreasing
resolution.  This is expected given that the EW of the absorption
line is conserved during the convolution, but the velocity
spread increases.  Lowering the resolution affects the \dind\ of individual
absorbers differently, depending on their kinematic profiles.
For example, the success of the \dind\ in our tests actually improves
for FWHM = 0.4 \AA, but this is just an artefact of the
particular absorbers in our sample.  The results of our convolution test 
highlight the need for a larger sample to statistically test
the effect of resolution on the \dind\ cut-off.  Based on our current data,
the recommended values of $D$ as a function of the
resolution, as well as the KS probabilities, are given in Table
\ref{convolve_table}.  The fraction of absorbers above the
recommended $D$ cut-off which are DLAs is given in the column
labelled `Success rate'.  The fraction of DLAs whose \dind\ is
below the recommended cut-off is given in the column labelled
`Missed DLAs'.  At  resolutions below 1.5 \AA\ it is still possible to
set a \dind\ that has a high success rate, but the fraction
of missed DLAs exceeds 15\% by FWHM $\sim$ 1.8\AA\ and the KS probability
indicates that the $D$-indices of the DLAs
and sub-DLAs are no longer significantly different.  Nonetheless,
a resolution of 1.5 \AA\  is within the reach of
many existing spectrographs, such as ISIS (William
Herschel Telescope), EFOSC (ESO 3.6-m telescope), FORS
(VLT) and ESI (Keck), as well as the very high resolution
instruments used for this study.  Unfortunately, however, based
on the resolution tests and the current data, the SDSS data
have slightly too low resolution for the \dind\ to be successfully
applied.

\begin{center}
\begin{table*}
\caption{Results of Resolution Tests}
\begin{tabular}{ccccc}
\hline
Resolution & Recommended &  Success &  Missed & KS \\
(\AA\ FWHM) & $D$-index & Rate (\%) & DLAs (\%)  & Prob. \\ \hline
0.4 & 5.7 & 91 & 0 & 0.4 \\
0.6 & 5.4 & 90 & 5 & 0.8 \\ 
0.8 & 5.1 & 90 & 5 & 0.8 \\
1.0 & 4.9 & 90 & 5 & 0.8 \\
1.2 & 4.6 & 90 & 5 & 0.8 \\
1.4 & 4.4 & 90 & 5 & 0.8 \\
1.6 & 4.2 & 90 & 10 & 0.8 \\
1.8 & 4.0 & 89 & 16 & 2 \\
2.0 & 3.9 & 89 & 16 & 3 \\
2.2 & 3.8 & 84 & 16 & 11 \\
\hline
\hline
\end{tabular}\label{convolve_table}
\end{table*}
\end{center}


\begin{figure}
\centerline{\rotatebox{270}{\resizebox{6cm}{!}
{\includegraphics{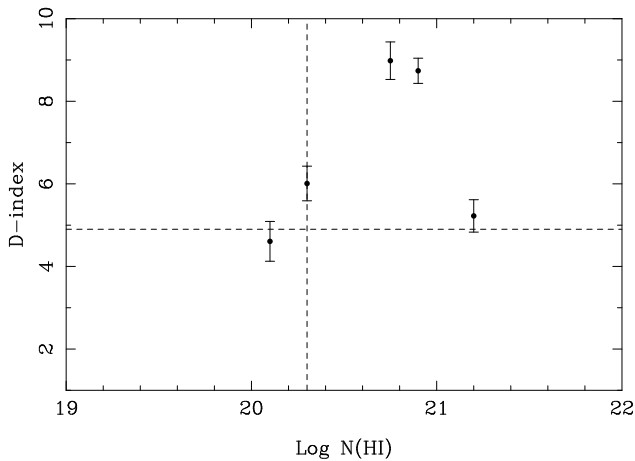}}}}
\caption{\label{ESI} $D$-indices measured from ESI spectra with 1 \AA\
resolution.  The horizontal dashed line shows the recommended
cut-off for this resolution from Table \ref{convolve_table}.}
\end{figure}


As a check of the convolution tests, we measured the $D$-indices
of five absorbers with known \nhi\ and existing ESI spectra provided
by J. Prochaska.  The absorbers are those towards
SDSS1027+1405 ($z_{\rm abs} =$ 2.441, \nhi\ = 20.30), SDSS2036-0555
($z_{\rm abs} =$ 2.280, \nhi\ = 21.20), SDSS1042+0117 
($z_{\rm abs} =$ 2.267, \nhi\ = 20.75), SDSS2059-0529
($z_{\rm abs} =$ 2.210, \nhi\ = 20.90) and Q1337+11
($z_{\rm abs} =$ 2.142, \nhi\ = 20.1).
The data were obtained with a 0.5 arcsec slit and
have a spectral resolution $\sim$ 1 \AA, so we adopt a
\dind\ cut-off of 4.9 from Table \ref{convolve_table}.
We plot the $D$-indices versus \nhi\ in Figure \ref{ESI},
where we see that the DLAs (4/5 absorbers) all have values
above the cut-off value whereas the one sub-DLA has $D < 4.9$.
Although more extensive testing is required, these initial results
from lower resolution spectra are encouraging.

Finally, since resolution has a significant impact on the measured \dind,
we have checked that the distribution of values in Figure \ref{dindex}
is not due to observing coincidence.  We checked that both the
distribution of absorber redshifts and spectral resolutions are
random with respect to \dind.    Although the spectral resolutions
listed in Table \ref{archive} do show some slight variation, there
is no anti-correlation between \dind\ and resolution. In fact,
the handful of absorbers
with FWHM $\ge$ 0.2 \AA\ all have high $D$-indices. The \dind\
of DLAs is high for all resolutions represented by the archival
sample, ranging from 0.11 -- 0.23 \AA.    This is
probably due to small numbers since a lower resolution will lead
to a lower \dind\ (see Figure \ref{convolve}).

\subsection{Signal-to-Noise}

We can do a similar test for the requisite signal-to-noise ratio
(S/N).  For two of the absorbers in our sample (one with a low
\dind, Q1247+267, and one with a high \dind, Q0058+019) we produce
a noiseless spectrum from a Voigt profile decomposition of the
Mg~II $\lambda$ 2796 line.  Many of the components are saturated,
so this `fit' does not yield accurate column densities and may
be degenerate in the absence of a simultaneous fit with lower
$f$-value species.  However, it does yield an accurate spectral
profile to which noise can be added artificially.  We add noise
from a Gaussian distribution and simultaneously create the
corresonding sigma error array which is used by the \dind\
algorithm in determining the velocity spread of the line.

We find very little change in the \dind\ as a function of S/N
in the range $10 <$ S/N $<100$, variations are $\le$ 2\%
for both the high and low \dind\ absorbers that we test.
Only at S/N $ < $5 do we see a significant change in the \dind,
although the magnitude of the change, and its direction
depend upon the structure of the absorber.  Nonetheless, it
is encouraging that the \dind\ technique is applicable for
spectra of relatively low quality.

\section{Discussion and Conclusions}

We have presented a new technique for discriminating between low
and high \nhi\ absorbers based solely on high resolution
spectroscopy of the Mg~II $\lambda$ 2796 line.  Qualitatively,
DLAs tend to be characterised by broad, heavily saturated Mg~II,
whereas sub-DLAs often have considerable residual flux over
their extent.  Quantitatively, we define the \dind, the ratio
of Mg~II $\lambda$ 2796 EW to velocity spread.  DLAs tend to have
high $D$-indices, with a cut of $D > 6.3 $ yielding a success
rate of 90\%.  
Previously, identification of DLA candidates based on EW cuts from
low resolution spectra had a DLA confirmation rate of only 35\%.
A larger sample of absorbers which is
more representative of the column density distribution function
(i.e. with a higher fraction of sub-DLAs than DLAs), may have
a lower success rate depending on the performance of the \dind\
at low \nhi.  This is difficult to judge from the current sample
since only 8 of our absorbers have \nhi\ $<$ 20.3.  Both of the
sub-DLAs with high $D$ can be marked out as `unusual' in some
way; Q0058+019 is known to have a very small impact parameter
to its parent galaxy and is rare in exhibiting solar
metallicity (Pettini et al. 2000) and Q0957+561B is a lensed QSO.  
Whilst we do not
suggest these factors as exemptions or explanations, it once
again highlights the need for better sub-DLA statistics.  
In addition, all DLAs with \nhi\ $>$ 20.5
in our pilot sample have $D > 6.3$, so that pre-selection 
with the \dind\ includes the systems which dominate the neutral
gas content.  Although the column density distribution function
evolves as a function of redshift (e.g. Prochaska, Herbert-Fort \&
Wolfe 2005; Rao, Turnshek \& Nestor 2006; Zwaan et al. 2005) it can 
be approximated to single power law $f(N) \propto N^{\alpha}$.
For an index $\alpha = - 1.5$, less than 10\% of the HI gas is in
systems with log \nhi\ $\le$ 20.5.  Therefore, although using
the \dind\ may miss some DLAs with lower column densities,
for the purposes of pre-selecting DLAs to
estimate the total neutral gas in DLAs, it is quite robust.  Similarly,
for any property that is weighted by \nhi, such as is common
practice with metallicity (e.g. Pettini et al. 1999; Kulkarni
et al 2005), the DLAs missed by \dind\ pre-selection will
have a relatively minor effect.

In order for the \dind\ to be a successful discriminant of DLAs
from sub-DLAs, a spectral resolution of at least 1.5 \AA\ FWHM
(at the wavelength of Mg~II $\lambda$ 2796)
is required.  Based on simulations which degrade the spectral
resolution of our archival data, we have tabulated the \dind\ which
yields the highest DLA return rate for a given instrumental FWHM.

The distinction between DLAs and sub-DLAs encapsulated by the \dind\
is one of spectral morphology.
The qualitative tendency for sub-DLAs to have more extended
velocities for their EWs with a larger number of distinct components
was previously noted by Churchill et al. (2000). Rao \& Turnshek (2000)
also discussed the difference in spectral morphology between high
and low \nhi\ absorbers noting that `the DLA 
systems generally appear to exhibit somewhat greater saturation.' 
The physical
reason for this difference, which leads to the difference in
$D$-indices in the two populations, is still unclear, but we speculate
that it may be driven by impact parameter (see also Rao \& Turnshek 2000).  
In their study of
a gravitationally lensed QSO, Ellison et al. (2004b) found that
the coherence scale of weak (EW $<$ 0.3 \AA) Mg~II absorbers is 
$\sim$ 2 \hkpc\ ($\Omega_{\Lambda}$ = 0.7, $\Omega_{\rm M}$ = 0.3), 
whilst stronger absorbers have much larger sizes.
Ellison et al. discussed a scenario in which galactic halos were
composed of numerous Mg~II absorbing `clouds' whose filling factor
was larger near the centres of galaxies.  In this picture, a sightline
intersecting at small impact parameters would show absorption from
many Mg~II regions, but within a relatively narrow velocity range.  
Conversely, sightlines at larger impact parameters would exhibit more 
disjointed absorption complexes and may have much wider velocity
spreads (depending on the kinematics of the individual galaxy).
This would  lead to an anti-correlation between \dind\ and impact
parameter.  We can test this prediction with the galaxy--Mg~II
absorber sample of Churchill et al. (2000) for which galaxies
have been identified for 16 Mg~II absorbers (albeit with EWs
typically lower than those considered in this paper).
Although Churchill et al. (2000) define velocity spread differently
from us, the ratio with EW (the analog to our \dind) does
anti-correlate with impact parameter.
If the HI column density also falls off with impact parameter,
this would lead to a correlation between \dind\ and \nhi\ and
qualitatively explain the distribution of points in Figure
\ref{dindex}.

We conclude that, based on these preliminary investigations,
a combination of EW and velocity spread may be a powerful
way to pre-select DLA candidates. Although more data are required
to test the robustness and selection biases of this technique
it could potentially increase the yield of pre-selected DLA
candidates by a factor of 2--3 over current methods.
With more data available, it will also be possible to extend the
investigation of the \dind\ to other transitions which may
be suitable for DLA identification, such as Fe~II
transitions longwards of $\sim$ 2300 \AA.

\section*{Acknowledgments}

I am indebted to those people who generously shared segments of their
reduced echelle spectra for the purposes of this study and my
collaborators on
previous projects for which these data were collected:  
Ramana Athreya, Chris Churchill, Mirka Dessauges-Zavadsky, Nissim
Kanekar, Sebastian Lopez, C\'eline P\'eroux, Max Pettini,
Jason X. Prochaska and Chuck Steidel.

\end{document}